\documentclass{PoS}

\newcommand \ra  {\rightarrow}

\newcommand \A {\alpha}

\newcommand \bvec{\left( \begin{array}{c} }
\newcommand \evec{\end{array} \right)}

\newcommand \bea{\begin{eqnarray} }
\newcommand \eea{\end{eqnarray} } 

\newcommand {\be} {\begin{equation}}
\newcommand {\ee} {\end{equation}}

\newcommand {\mbx} {\mbox{}}

\title{Effects of drag induced radiation and multi-stage evolution on heavy quark energy loss}

\ShortTitle{Heavy quark drag induced radiation}

\author{\speaker{Abhijit Majumder}, Shanshan Cao, Chun Shen,\\
        Department of Physics and Astronomy, Wayne State University, Detroit, Michigan 48201, USA.\\
        E-mail: \email{majumder@wayne.edu}}
        		
	\author{Guang-You Qin\\
	Institute of Particle Physics and Key Laboratory of Quark and Lepton Physics (MOE), Central China Normal University, Wuhan, China.}
	
\abstract{Heavy quarks serve as ideal probes of the QGP properties produced in energetic nuclear collisions, and provide a unique opportunity to study the mass effects on parton energy loss. We develop a multi-stage approach for heavy quark evolution inside the QGP, in which heavy quarks first undergo a rare-scattering multiple-emission evolution at momenta large compared to their mass (sensitive only to the transverse diffusion coefficient $\hat{q}$ ), and then evolve through a single-scattering induced emission (Gunion-Bertsch) stage at momenta comparable to their mass [sensitive to not only $\hat{q}$, but also the longitudinal drag $\hat{e}$ and diffusion $\hat{e}_2$ coefficients]. This multi-stage approach is coupled to a (2+1)-D viscous hydrodynamic model for a quantitative investigation of charm vs. beauty quark energy loss inside the QGP. Based on this approach, we find that drag induced radiation has a considerable impact on the energy loss of intermediate $p_T$ massive beauty quarks. This effect increases the suppression of B mesons and narrows the difference between the $R_{AA}$ of B and D mesons. Our results are consistent with the experimental data at the LHC and contribute to a more quantitative understanding of the transverse momentum dependence of the mass hierarchy of parton energy loss inside the QGP.}

\FullConference{International Conference on Hard and Electromagnetic Probes of High-Energy Nuclear Collisions\\
		30 September - 5 October 2018\\
		Aix-Les-Bains, Savoie, France}

\begin{document}

\section{Introduction.}

The modification of the spectrum and substructure of jets, 
and heavy flavors have become leading probes of the internal structure of the Quark-Gluon-Plasma (QGP)~\cite{Majumder:2010qh}. 
Of these the modification of heavy flavor hadrons with an intermediate momentum, of the order of, and larger than the mass of the heavy quark i.e. $p_T \geq M_Q$, are of special interest. 
Very high energy heavy quarks whose mass is negligible compared to their momentum are expected to behave essentially as light quarks. 
Bremsstrahlung from a heavy quark produced in a hard interaction are known to be effected by the dead cone effect~\cite{Dokshitzer:2001zm}, where collinear radiation is suppressed due to the presence of mass. This should only become a noticeable effect when the momentum $p$ is not much larger than $M_Q$. As a result, one naively expects that there will be a mass dependent hierarchy in the energy loss experienced by different parton flavors.

The above expectation is based on the assumption that the mechanism of energy loss in both light and heavy flavors is similar, and depends on the same transport coefficients. In these proceedings, we report on ongoing research which indicates that there are systematic differences between light and heavy flavor \emph{radiative} energy loss. In the case of heavy quarks with momentum $p \geq M_Q$, the drag and longitudinal coefficients $\hat{e}$ and $\hat{e}_2$ cause a non-negligible contribution to the \emph{radiative} energy loss. 
This effect is beyond the additive contribution of energy lost by drag experienced by the heavy quark in the medium, also dependent on the same coefficient $\hat{e}$. 
This increase in radiative loss sourced by $\hat{e}$ and $\hat{e}_2$ grows with the mass of the parton. Also, it is now widely accepted that parton showers go through several dynamically different phases~\cite{Majumder:2015qqa}, a vacuum like emission phase where the parton engenders multiple emissions with few scatterings, followed by multiple scattering induced rare emission phase (BDMPS phase)~\cite{Baier:1996sk}. This may also involve a strong coupling phase~\cite{Casalderrey-Solana:2014bpa} when both the energy and virtuality are comparable to 1~GeV. 
In contrast, for the case of heavy flavors, intermediate energy heavy quarks do not possess a 
multiple scattering induced rare emission phase, but rather a Gunion-Bertsch like few scattering per emission phase. As a result there are several differences between light and heavy flavor energy loss. 

\section{Drag induced radiation.}
\label{DR}

Consider the production of a high energy quark (momentum $p \gg M_Q$ the mass of the quark) in the deep-inelastic scattering of a virtual photon on a large nucleus. The momentum components of the quark are given as, 
\bea
p_q = \left[ p^+, p^-, p_\perp\right] \sim \left[  \lambda , 1, \sqrt{\lambda} \right]Q, \label{HardLightParton}
\eea
where, $Q$ is the hard scale of the process and $\lambda \ll 1$ is a dimensionless scaling variable.  As is typical for DIS, the outgoing quark propagates in the negative $z$ direction with a large $p^- \sim Q$. The virtuality of this parton is $p^2 = \lambda Q^2$. As a result, the radiated gluon has a transverse momentum $l_\perp \sim \sqrt{\lambda} Q$, and this causes the exchanged gluon's transverse momentum $k_\perp $ to also scale as 
$\sqrt{\lambda} Q$. Calculations involve an expansion in $\lambda$ with terms of order higher than $\lambda$ being ignored. The results of this expansion could be converted into the standard SCET$_{\rm G}$ results by simply replacing $\sqrt{\lambda} \ra \lambda$. 
This choice of scaling variable, i.e., $\sqrt{\lambda}$ instead of $\lambda$ is dictated by the need to encompass, high virtuality light flavors, low virtuality light flavors and heavy flavors in a single formalism. 

Within this formalism, it can be demonstrated that radiative energy loss for light flavors is dominantly controlled by the transverse diffusion coefficient $\hat{q}$, with the effect of $\hat{e}$ and $\hat{e}_2$ being suppressed in a $\lambda$ expansion. The final expression for the case of single gluon emission spectrum that is stimulated by scattering is given as, 
\bea
\frac{dN_g}{dy d l_\perp^2} = \frac{\A_S C_F}{2\pi l_\perp^4} P(y) \int_0^{L^-} d\zeta^- \hat{q}(\zeta^-) \left[ 2\sin^2\left( \frac{l_\perp^2}{4 p^- y (1-y)} \zeta^-\right)  \right].   \label{GW}
\eea
In the equation above, the variables $y, l_\perp$ represent the longitudinal momentum fraction and transverse momentum of the radiated gluon, which is produced as a result of the scattering at the longitudinal location $\zeta^-$ which lies between the origin where the hard quark enters the medium at $\zeta^- = 0$ to where it exits the medium at $\zeta^- = L^-$. While this is not evident in a semi-analytical (or event averaged) calculation, in Monte Carlo simulations, the location of the emission of the radiated gluon rarely exceeds the formation length $\tau^- = 2 p^-y (1-y)/ l_\perp^2 $.

High energy heavy quarks, whose energy and longitudinal momentum is orders of magnitude larger than the mass of the quark, formed in hard scatterings, tend to have the same 
momentum scaling relations as light quarks [Eq.~(\ref{HardLightParton})], shower gluons and loose energy in much the same way.   
As the energy of the heavy quark depletes and its longitudinal momentum and energy become comparable to its mass, it enters a different phase. For such intermediate energy heavy quarks, an analysis of the same diagrams that led to Eq.~(\ref{GW}) above, using a heavy quark mass $M \sim \lambda Q$, and momentum components that scale as 
\bea
p_Q \sim \left[ \sqrt{\lambda} , \sqrt{\lambda},  \lambda \right] Q,
\eea
one obtains a  modified single gluon emission spectrum which depends not only on $\hat{q}$, but also on the longitudinal drag $\hat{e}$ and diffusion $\hat{e}_2$ coefficients~\cite{Abir:2015hta,Abir:2014sxa}:
\bea
\frac{dN^Q_g}{dy d l_\perp^2} =  \frac{ \A_S C_F}{2\pi} 
P(y) \!\!\!\!\int d\zeta^- \frac{\left[ \hat{q} \left\{  1 - \frac{y}{2} - \frac{y^2 M^2}{l_\perp^2}\right\} + \hat{e} \frac{y^2 M^2}{l^-} + \hat{e}_2 \frac{y^2 M^2}{2(l^-)^2}\right]}{\left( l_\perp^2 + y^2 M^2 \right)^2} 2 \sin^2\left( \frac{l_\perp^2 + y^2 M^2}{ 4 l^- y(1-y)} \zeta^-\right)\!. \label{AM}
\eea
The reader will note that Eq.~(\ref{AM}) reduces to Eq.~(\ref{GW}) when $M \ra 0$. Even though not mentioned in the equation above, all three diffusion coefficients ($\hat{q}, \hat{e}. \hat{e}_2$) are functions of location $\zeta^-$. The range of the $\zeta^-$ integral is not from the origin, as this type of emission takes place after the heavy quark has already undergone a DGLAP like phase. As such we have not provided limits on the $\zeta^-$ integral. How the start and end location is determined is discussed in the next section. As demonstrated in Ref.~\cite{Cao:2017crw}, the radiated gluon has momentum components $l \sim [\lambda, \lambda, \lambda] Q$, and thus $y \sim \sqrt{\lambda}$. 
As a result, the ratio $yM/l_\perp \sim 1$. Similar terms, which depend on $\hat{e}, \hat{e}_2$ are parametrically suppressed in the case of massless flavors, and add extra contribution to the radiative loss of an intermediate energy heavy quark. While the longitudinal drag and diffusion coefficient lead to elastic energy loss of the heavy quark, they can also change the virtuality of the quark and thus lead to heavy quark radiative loss. 

\begin{figure}[h!]
$\mbx$
\includegraphics[width=0.40\textwidth]{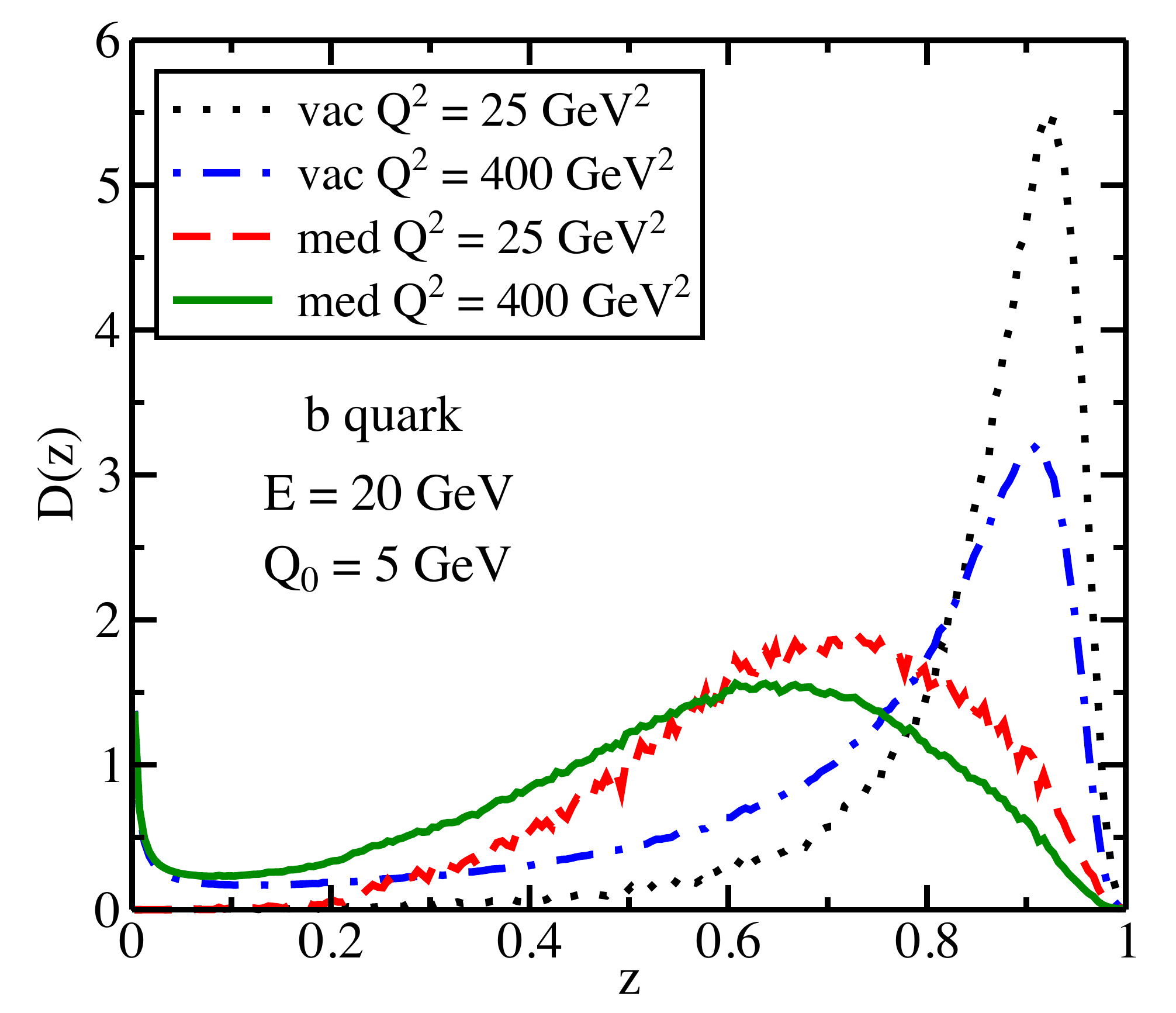} 
\hspace{1cm}
\includegraphics[width=0.47\textwidth]{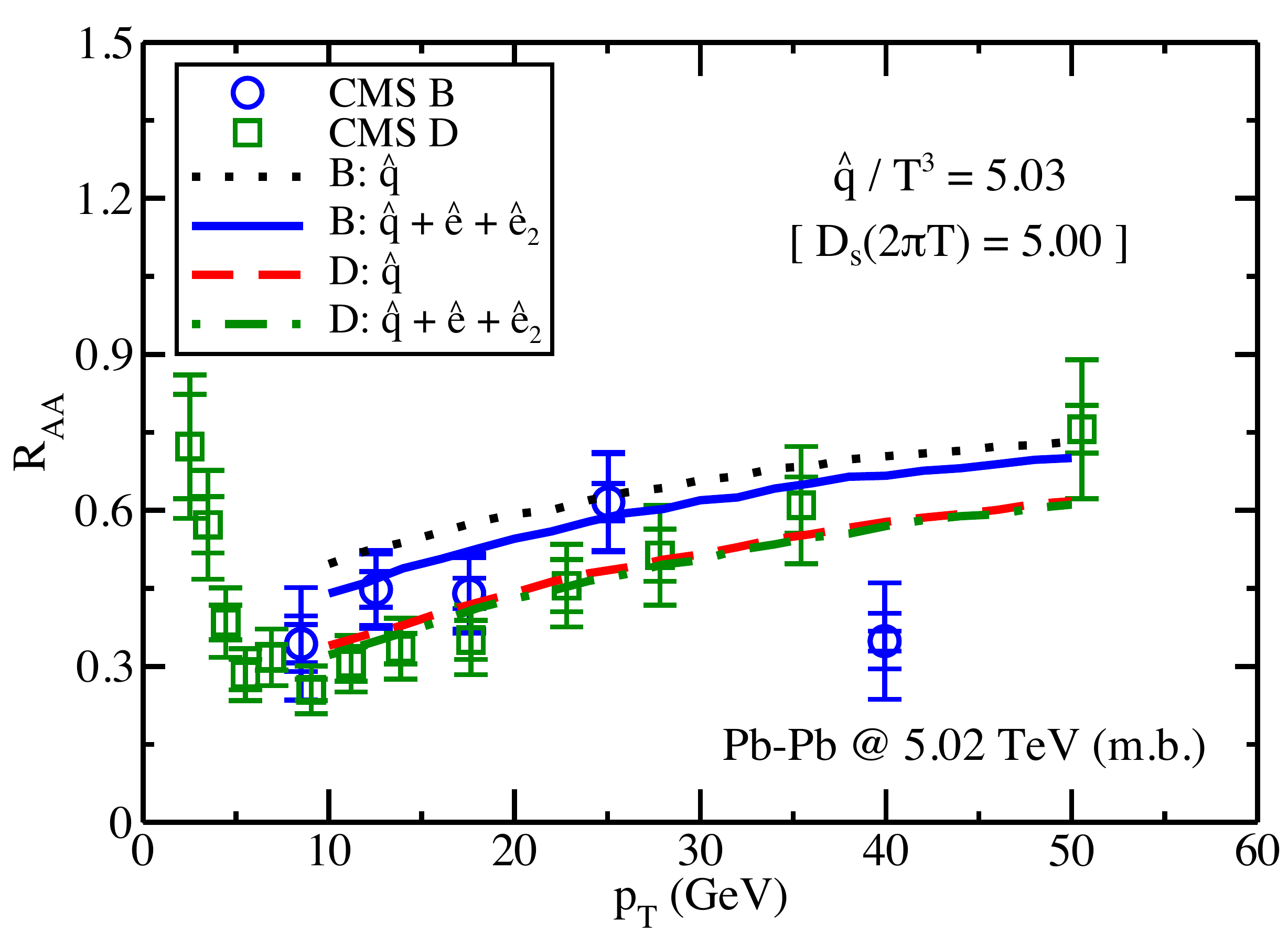} 
    \caption{Left: The b to B fragmentation function as measured at $Q_0=5$~GeV in vacuum and then evolved via Eq.~\ref{AM}, to a medium modified fragmentation function at the scale $Q_0=5$~GeV, then further modified via the medium modified DGLAP equation to the hard scale $Q=20$~GeV. This is compared with vacuum evolution from from $Q_0$ to $Q$. 
    Right: The nuclear modification factor for D and B mesons as measured by the CMS collaboration. The lines are theoretical 
    calculations based on the inclusion of gluon radiation induced by drag and longitudinal diffusion, in addition to transverse diffusion.}
    \label{plot}
\end{figure}

\section{Multi-stage Evolution, Results and discussion.}
\label{ME}

Hard partons produced in hard interactions, start off with virtuality scales that are much larger than any medium scale and as such engender vacuum like radiation. As the virtuality depletes, the partons enter the in-medium high energy high virtuality regime, where the momentum components of the light parton scale as in Eq.~(\ref{HardLightParton}). In this phase, it undergoes vacuum like emissions with a minor perturbative modification from scattering in the medium, this is the radiation dominated phase or DGLAP~\cite{Dokshitzer:1977sg,Gribov:1972ri,Altarelli:1977zs} like phase. In this phase the formation time of a typical radiation is given as 
$\tau \sim 1/(\lambda Q)$, which is rather short and thus reduces the amount of scattering felt engendered by the parton. 
In this regime, as well as in the vacuum regime that precedes it, both heavy and light quarks behave in a similar fashion as the virtuality scale far exceeds both mass and medium scales.

As the virtuality depletes, and approaches the medium scale of $\hat{q}\tau$ ($\tau$ is the formation time of a given radiated gluon), light partons enter the BDMPS phase where partons engender multiple scattering per emitted gluon. Heavy flavors however, due to the dead cone effect, enter a Gunion-Bertsch (GB) like phase with a few scatterings per emission. The predominant momentum scale of the emitted gluon is $l \sim (\lambda, \lambda, \lambda) Q$, i.e., at large angle. These gluons have a short formation time and thus encounter very few scatterings. As such, a full calculation of heavy quark energy loss will involve both a DGLAP like phase coupled with a GB phase. The transition from the DGLAP to the GB phase is set to occur at the scale $Q_0$. In these proceedings, $Q_0$ is chosen as the heavy meson mass of 5~GeV.  As a result, one should understand that 
DGLAP phase to be active from $\zeta=0$ to $\zeta \simeq2E/Q_0^2$, and the GB phase from this distance to exit.

Multi-scale calculations are presented in Fig.~\ref{plot}. The left panel represents calculations for the medium modified fragmentation function, from different stages of the multi-stage evolution. The vacuum fragmentation functions are measured from PYTHIA at the input scale of $Q\sim 5$~GeV, i.e., the mass of the B meson (dotted line). This is then evolved in vacuum up to a hard scale of $20$~GeV (dot-dashed line). Also shown is the effect of the Gunion-Bertsch evolution on the input fragmentation function (dashed line), and medium modified evolution of this up to the hard scale of $Q=20$~GeV. These calculations are carried out in a (2+1)-D viscous hydrodynamic medium which has been calibrated to yield the experimental spectrum and $v_2$ of low momentum hadrons in $Pb$-$Pb$ collisions at $\sqrt{s_{\rm NN}} = 5.02$~TeV. The minimum bias nuclear modification factor calculated with these medium modified fragmentation functions is plotted in the right panel of Fig.~\ref{plot}, where the effect of the different transport coefficients on the $R_{AA}$ can be clearly seen. Scaling $\hat{e}$ and $\hat{e}_2$ as in Ref.~\cite{Qin:2009gw}, and using the known value of $\hat{q}$ as measured by the JET collaboration~\cite{Burke:2013yra} one obtains a very good fit with the experimental data. Also interesting to note is the effect of the larger mass of the B quark on the extra suppression from the $\hat{e}$ and $\hat{e}_2$ terms from Eq.~(\ref{AM}).

\section{Acknowledgements}
The work of C.S, S. C. and A. M. is supported in part by the U.S. Department of Energy (DOE) under grant number 
\rm{DE-SC0013460} and in part by the National Science Foundation (NSF) under grant number \rm{ACI-1550300}. C. S. acknowledges support from a RIKEN fellowship. 
G.-Y. Q. is supported by the NSFC of China, under grant numbers 11775095, 11890711, and 11375072.

\bibliographystyle{JHEP}
\bibliography{refs}

\end{document}